# End-To-End Audiovisual Feature Fusion for Active Speaker Detection

Fiseha B. Tesema[a*], Zheyuan Lin[a], Shiqiang Zhu[a, d], Wei Song[a*], Jason Gu[a, b], Hong Wu[c]

[a]Interdisciplinary Innovation Research Institute, Zhejiang Lab, Kechuang Avenue, Zhongtai Sub-District, Yuhang District; [b]Electrical and Computer Engineering, Dalhousie University, NS B3J 2X4, Halifax, Canada; [c]School of Computer Science and Engineering, University of Electronic Science and Technology of China, Shahe Campus:No.4, Section 2, North Jianshe Road, Chengdu, China; [d]Ocean College, Zhejiang University 866 Yuhangtang Rd, Xihu, Hangzhou, China

## ABSTRACT

Active speaker detection plays a vital role in human-machine interaction. Recently, a few end-to-end audiovisual frameworks emerged. However, these models' inference time was not explored and are not applicable for real-time applications due to their complexity and large input size. In addition, they explored a similar feature extraction strategy that employs the ConvNet on audio and visual inputs. This work presents a novel two-stream end-to-end framework fusing features extracted from images via VGG-M with raw Mel Frequency Cepstrum Coefficients features extracted from the audio waveform. The network has two BiGRU layers attached to each stream to handle each stream's temporal dynamic before fusion. After fusion, one BiGRU layer is attached to model the joint temporal dynamics. The experiment result on the AVA-ActiveSpeaker dataset indicates that our new feature extraction strategy shows more robustness to noisy signals and better inference time than models that employed ConvNet on both modalities. The proposed model predicts within 44.41 ms, which is fast enough for real-time applications. Our best-performing model attained 88.929% accuracy, nearly the same detection result as state-of-the-art work.

**Keywords:** Audiovisual active speaker detection, MFCC, VGG-M, Audiovisual fusion, BiGRU

## 1. INTRODUCTION

Exploiting audiovisual analysis of video speech contents has received significant attention due to its practical application in several areas and academic values. It has been widely studied in several areas, such as visual speech recognition, lip-reading, and audiovisual speech separation and enhancement. However, it has been less explored in active speaker detection (ASD) works. ASD aims to detect speaking persons from visible human instances in the video at any given time. It plays a significant role in the preprocessing steps for visual speech recognition and other numerous applications such as video annotation [1], human-robot interactions, speaker diarization or re-framing video[2], speech transcription, and speech enhancement [3,4].

Several ASD works that employed deep architectures (combining ConvNets and recurrent neural networks) to learn active speakers from audiovisual inputs were introduced after the publicly published AVA-Activespeaker detection dataset [2]. The vast majority follows a two-step approach where features were first extracted from audio and visual modalities with ConvNet, fused, and then fed to a classifier. Joseph et al. [2] proposed an end-to-end multimodal active speaker detection framework with a two-stream convolution network for audiovisual feature extraction followed by a recurrent neural network for classification. Chong and Kazuhito [5] proposed two end-to-end active speaker detection frameworks which leverage an encoder-decoder-based prediction network. The visual-coupled-embedding network integrates the features extracted from the raw image and optical flow with a single network, and the independent-embedding network uses separate networks to extract features from optical flow and raw image. Both [2] and [5] used a MobileNet [6] to extract features from both modalities. Joon [7] introduced a two-stream VGG-M-based audiovisual encoding framework attaching the Bidirectional Long Short Term Memory (BLSTM) layer for each modality. They fused the outputs of each encoder to predict with a fully connected (Fc) layer. Yuan-Hang [8] introduced an audiovisual framework built using the 3D-ResNet18 visual model pertained for audio-to-video synchronization tasks. Juan et al. [9] proposed a multistage framework that leverages a two-stream-based architecture that models the relationships between multiple speakers over a long time horizon. However, these works followed similar feature extraction strategies that extract features from the visual and audio stream with ConvNet and neglected to explore the direct fusing of MFCC features with visual ConvNet features for active speaker detection. In addition, these works emphasized improving the accuracy of their model by proposing complex networks with larger input sizes, which is not suitable for real-time application and have not reported the inference time of their proposed model. Stefanov et al. [10] tried to address the real-time issue by proposing a Long Short Term Memory (LSTM) based prediction model that combines visual features vector with audio labels. The visual feature vectors are extracted from the input image using AlexNet, and the labels are generated from acoustic signal inputs using the interval detected by Voice Activity Detection (VAD) module. However, it is a speaker-dependent, leveraged old backbone network to extract features (AlexNet), designed for a limited number of speakers and trained on the small dataset recorded in a confined space. This work aims to explore ASD, proposing a less complex framework that could be used for real-time application by directly fusing the MFCC audio features with CNN channels.

*Corresponding author: fisehab@zhejianglab.com

Mel Frequency Cepstrum Coefficients (MFCC) [11] is used as a baseline in the feature extraction method, which aims to improve the accuracy of the speech activity detection applications. MFCC is one of the most broadly used approaches in audio recognition, and a wide range of different sounds like surveillance-related events [12], soundscapes [13], or even animal sounds [14] and numerous works related to events and anomalies detection in an outdoor environment. Similarly, recently MFCC has been leveraged in lip-reading and active speaker detection combining dense features extracted from MFCC with visual information using ConvNets [2,5,7].

This paper presents a novel end-to-end audiovisual framework that (see Figure 1) aims to detect an active speaker in real-time. It consists of the visual stream that extracts features directly from cropped facial images using VGG-M and the audio stream that extracts the MFCC features from audio waveforms. The video stream consists of a newly adopted VGG-M network followed by a 2-layer of Bidirectional Gated Recurrent Unit (BiGRU). The audio stream consists of the MFCC feature extractor followed by a 2-layer BiGRU. Each 2-layer BiGRU is used to model the temporal dynamics in each stream. Finally, the information of the different streams is fused and fed to 1-layer BiGRU, which models the joint temporal dynamics. To the best of our knowledge, the proposed model is the first end-to-end model to explore ASD proposing a model that performs a fusion of features extracted from raw facial regions using ConvNet with raw MFCC features. The paper has the following contributions:

1) Proposed a novel two-stream-based active speaker detection model combining the audio MFCC features with features extracted from an image with a VGG-M network. The network is trained end-to-end without the use of any pre-trained networks.

2) We explored three audio features extraction approaches, the MFCC, applying VGG-M on the top of MFCC, and applying two fully connected (Fc) layers on the top of MFCC. The ablation experiments show that the raw MFCC features-based model attained nearly the same detection accuracy as the model that employed a VGG-M on the top of the MFCC and attained better detection results than the model that employed two Fc layers on the top of MFCC.

3) We also investigated the proposed model's performance towards a noisy signal generated by applying Gaussian noise to the AVA-Active Speaker validation set. The experiment reveals that the raw audio MFCC-based model showed robustness.

4) We reported the inference time of the proposed models. It predicts within 44.41ms, which is applicable for real-time applications.

5) Our best-performing model attained 88.929% accuracy, nearly the same detection result as state-of-the-art work.

## 2. THE PROPOSED METHOD

This section discusses the network architecture of the proposed model. As shown in Figure 1, the network consists of two asymmetric streams for audio and video. The video stream extracts features from the raw cropped facial images with the 3D ConvNet VGG-M followed by two layers of BiGRU, each layer consisting of 256 units. The audio stream uses a direct MFCC feature extracted from the input audio waveform followed by two layers of BiGRU (each layer consists of 256 units). The two layers of BiGRU attached to each stream aim to model the temporal dynamics of the features in each stream. The outputs of each BiGRU from each stream are concatenated and then fed to another BiGRU of 512 units to model the joint temporal dynamics. The output layer is a softmax layer that provides a label for input frames. The entire model is trained end-to-end, enabling the joint learning of features and classifiers. An independent auxiliary classifier was added to each embedding modality with corresponding cross-entropy loss to boost the prediction network by using visual and audio embedding. The detailed audio and visual networks will be discussed as follows.

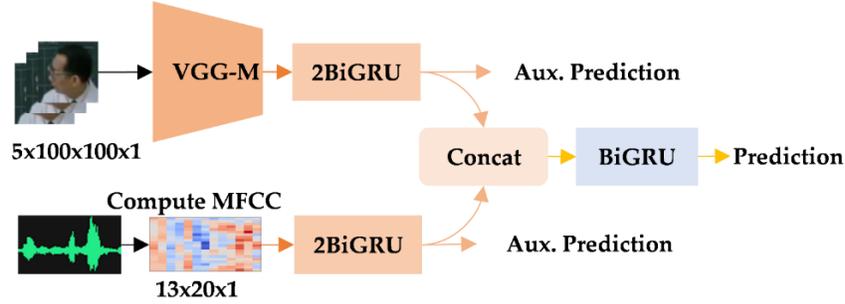

Figure 1. The proposed framework: Top, the VGG-M directly extracts features from the raw cropped facial image. Bottom, the MFCC features are extracted from raw audio waveforms. The output of each stream is fused and fed to BiGRU to model their temporal dynamics jointly. An independent auxiliary classifier was added to each embedding modality with corresponding cross-entropy loss.

**2.1 Video stream**

As shown in Figure 1, the input format to the visual network is a sequence of face regions. The network inputs five stacked grayscales cropped face regions of $100 \times 100$ dimensions at once, corresponding to 0.5 seconds. The newly adopted VGG-M architecture is based on that of [15]. It was first designed for visual speech recognition and later adopted for lip-reading [4] and active speaker detection tasks [7]. As shown in Figure 2 (a), unlike the vanilla VGG-M that uses $7 \times 7$ filters, we adopted the VGG-M network that uses filter size of $5 \times 7 \times 7$ [16] to capture the motion information across five frames.

**2.2 Audio stream**

For the audio stream, we explored three possibilities in this work, (1) adopting the VGG-M, (2) direct use of MFCC, and (3) two Fc layers. MFCC features were extracted using a 25ms analysis window with a stride of 10ms, yielding 100 audio frames every second. The size of the MFCC feature used in these audio streams is $13 \times 20 \times 1$, 20 frames in time-direction and 13 cepstral coefficients in other direction.

**MFCC:** Human auditory ability to sound signals is not linear. Mel-frequency Cepstral Coefficients (MFCC) is a feature extraction method based on human hearing characteristics. MFCC [11,17] is a filter banks-based cepstral-domain features extraction method. After applying the Fast Fourier Transform (FFT) to a given windowed signal, a Mel-Scaled filter bank (based on a nonlinear frequency scale inspired by physiological evidence of how the human perception of speech signals works) is used. This triangular filter bank divides the spectrum non-linearly following the Mel scale, where the lower frequency filters have smaller bandwidth than the higher frequency ones. The Mel scale follows a linear frequency spacing below 1 kHz and a logarithmic spacing above 1 kHz. The number of filters and the chosen frequency range must be previously defined. The last stage consists of ranging the coefficients according to their significance, which is accomplished by calculating the Discrete Cosine Transform (DCT) of the logarithmic outputs from the filter bank. MFCC is a very popular feature extraction technique in speech recognition because it has auditory methods such as human perception concerning frequency [18].

**VGG-M:** is the most widely adopted network for audio feature extraction in lip-reading [4,15], and active speaker detection [7]. The adopted VGG-M network filter size is modified for the audio input sizes. The detailed illustration of the layers is shown in Figure 2 (b).

**Two Fc layers:** following the lip reading work [19] that extracts features from spectrogram with two layers of a fully connected network. We explored our model detection performance, leveraging the 2fc layer on the top of the MFCC features feature. The adopted 2 Fc layers are shown in Figure 2 (c).

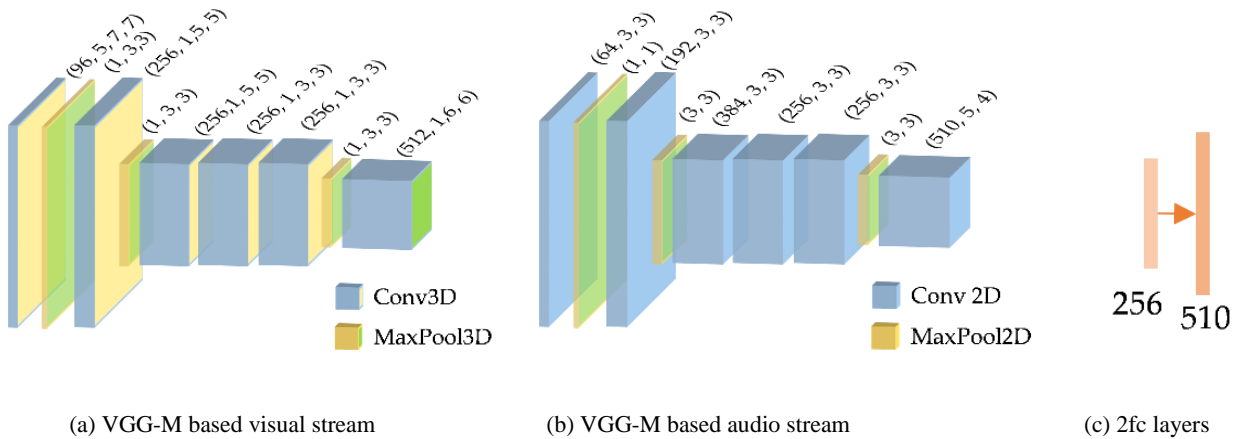

| (a) VGG-M based visual stream | (b) VGG-M based audio stream | (c) 2fc layers |

Figure 2. The adopted visual and audio streams.

## 3. EXPERIMENTS

This section describes the dataset, prepossessing, implementation details, adopted loss function, and evaluation metrics.

### 3.1 Dataset

The AVA-Activespeaker detection dataset was used to evaluate our method. The dataset contains temporally labeled face tracks in video, where each face instance is labeled as speaking or not. Compared with the Columbia [20] and VoxCeleb [21] datasets, the AVA-ActiveSpeaker dataset has much more labeled data, including 3.65 million face frames (38.5 hours) and 113 training videos (28,108 face tracks), and 32 testing videos (7,900 face tracks). In addition, the AVA-ActiveSpeaker dataset is very challenging due to low resolution (e.g., people in the distance) or occlusion (e.g., profile faces). 44.6% of labeled faces are smaller than 100 pixels wide, and 48.2% are challenging to detect the face mouth region by the state-of-the-art face landmark detection library [5].

### 3.2 Prepossessing

Images were extracted from the video with a frame rate of 10fps. The frames are transformed to grayscale and normalized with respect to the overall mean and variance. On the other hand, the MFCC features are also normalized with respect to the overall mean and variance.

### 3.3 Implementation details

All the models in the experiments were trained under the same conditions. Our implementation is based on the Keras library [22] and trained on a single NVIDIA Titan RTX GPU of 24 GB memory. The network is trained using the SGD optimizer [23], with a 0.01 and 0.9 momentum learning rate. During training, the number of samples for positive and negative classes is balanced in each mini-batch to avoid bias in the training data. Early stopping with a delay of 5 epochs and a Batch-normalization were applied after each layer for audio and video streams to avoid overfitting. We also applied a dropout, a rate of 0.5%, after concatenation and before single layer BiGRU.

### 3.4 Loss function

We define the loss function $l(w)$ as a cross-entropy loss between the predictions $P_j$ and labels $y_j$:

$$l(w) = -\sum_j y_j \log(p_j) + \alpha \| w \|^2 \qquad (1)$$

where $\alpha$ is a regularization hyperparameter. Furthermore, to encourage the prediction network to make use of both the audio and visual embedding, we add independent auxiliary classification networks on each modality with corresponding cross-entropy loss. Our final loss is then a combination of all terms:

$$l(w) = l_{av} + \alpha_a L_a + \alpha_a L_v \qquad (2)$$

where $L_{av}$, $L_a$ and $L_v$ are the cross-entropy loss (Eq. (1)) of audiovisual, audio-only, and visual-only networks, and following settings in [2,5], we set $\alpha_a = \alpha_b = 0.4$ to place a lower weight on the individual modality performance.

### 3.5 Evaluation metrics

We adopted "area under the Receiver Operating Characteristics curve (AUC)" as a metric. To test the robustness of the proposed model against noise signals, we followed the evaluation setting in [10], which adds a stationary noise (white Gaussian Noise) to the audio signal. For every record, we generate noise by sampling it from Gaussian distribution using standard deviation $\delta$ and a root-mean-square of the signal (shown in equation (3)) with a zero-mean noise ($\mu_{noise} = 0$). Finally, the generated noise is added to every AVA-ActiveSpeaker validation set to generate a noisy validation set.

$$\delta_{noise} = \sqrt{\sum \frac{(n_i - \mu_{noise}{}^2)}{n}} \qquad (3)$$

## 4. EXPERIMENTAL RESULTS

Previous works heavily relied on VGG-M and Fc layers for audio feature extraction techniques. This section investigates the proposed model exploring three techniques of audio features, the raw MFCC feature (M1), features extracted by applying VGG-M on the top of the raw MFCC feature (M2), and features extracted by applying the 2-Fc layer on the top of raw MFCC (M3), listed in Table 1. The adopted VGG-M and two Fc layers audio stream networks are shown in Figure 2 a, b, and c. Further, we discuss the comparison of our work with other state-of-the-art works.

Table 1. Comparison of the proposed models' prediction time and accuracy.

| Model | Audio stream | Visual stream | BiGRU | Inf. time | Video Acu. | Audio Acu. | AV Acu. |
|---|---|---|---|---|---|---|---|
| M1 | MFCC | VGG-M | 2 | 44.41ms | 80.774 | 77.087 | 87.940 |
|    |      |       | 1 | 39.78ms | 79.651 | 76.736 | 87.480 |
| M2 | MFCC+2fc | VGG-M | 2 | 44.48ms | 80.577 | 77.250 | 86.847 |
|    |          |       | 1 | 40.37ms | 76.130 | 78.886 | 86.859 |
| M3 | MFCC + VGG-M | VGG-M | 2 | 47.44ms | 81.669 | 78.030 | 88.929 |
|    |              |       | 1 | 42.10ms | 77.856 | 79.191 | 88.069 |

### 4.1 Ablation Experiments

As shown in Table 1, before seating the BiGRU layers for each audio and visual stream to two layers, we undertake an ablation experiment on these models by seating a single BiGRU layer. All the experiments listed in Table 1 have been undertaken under the same parameter settings and the visual stream. M1 and M3 detection accuracy improved when we employed a two-layer BiGRU instead of a single BiGRU. M1 and M3 improved by 0.771% and 0.86%, respectively. The M2 does not show any performance gain. Instead, it less performed by 0.012% than the two layers BiGRU. After observing this, two layers of BiGRU were leveraged throughout the experiment.

On the other hand, as listed in Table 1, the result on the AVA-ActiveSpaker validation set shows that the M3 achieved 88.929% accuracy. However, the M1, without an additional feature extraction network, attained 87.940%, nearly the same detection accuracy result as the M3. The experiment indicates that applying the VGG-M to MFCC improved the detection result but did not bring significant accuracy. We infer from the experiment that raw MFCC features are robust enough to discriminate between human speech and non-speech signals without additional complications or leveraging additional ConvNet. On the contrary, the experiment reveals that leveraging Fc over the top of MFCC features degrades the model accuracy by 0.78%. Therefore, the experiments indicate that employing Fc layers on the top of the raw MFCC features did not help the model discriminate the speech signal.

To examine inference time and applicability for real-time application of the proposed models, we undertake experiments with the proposed models on the 11GB GeForce RTX2080Ti GPU. The inference time of the proposed models is listed in Table 1. M1 predicts within 44.41ms, whereas the M2 and M3 predict within 44.48ms and 47.44ms, respectively. These inference times prove the proposed model's applicability for real-time applications with limited computational resources.

On the other hand, to investigate the robustness of the proposed audiovisual methods to the noisy audio signal, we run experiments after generating the noisy audio data according to equation (3). The audiovisual and audio classifier detection result of all the models under noisy conditions are shown in Figure 3 (a) and (b). On all models, the audiovisual classifier and audio classifier are affected by the noisy audio inputs. The M3 audiovisual classifier is affected much by the noisy signal, and its accuracy drops from 88.929 to 86.584 or by 2.345%. Comparatively, M1 and M2 are less affected by a noisy signal; the accuracies are dropped by 1.739% and 1.383%, respectively. We inferred from these experiments that the raw MFCC features are more robust to noisy signals than the audio feature extracted by ConvNet. However, using two fully connected layers on the top of raw MFCC shows very slight robustness than the MFCC features. Similarly, as shown in Figure 3 (b), the M3 audio classifier is highly affected by the noise signal than M1 and M2, M1 drooped by 3.246%, M2 by 3.258%, and M2 drooped 3.655%. The detection accuracy for the video classifier is not reported because the noisy signal only affects the weights of the audio stream.

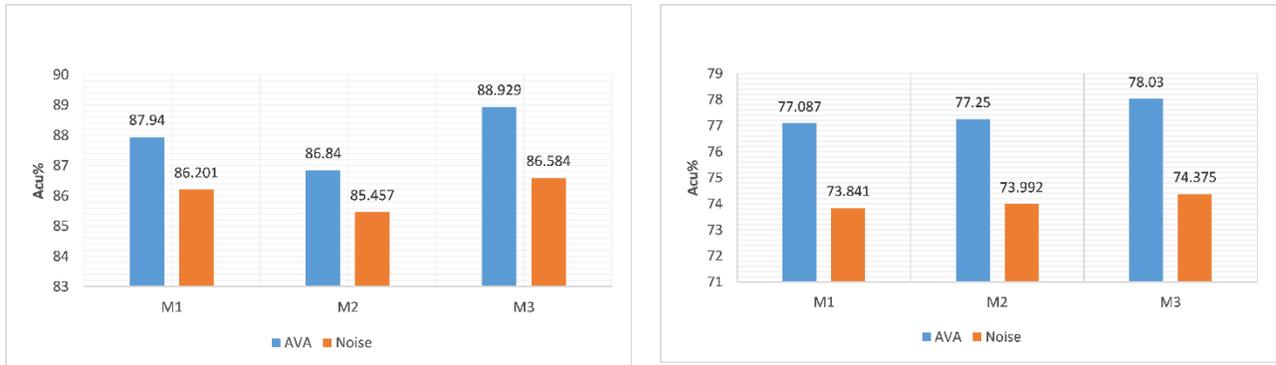

(a) Audiovisual detection accuracy on Noise and AVA-Activespeaker validation set (AVA) audio signal

(b) Audio detection accuracy on Noise and AVA-Activespeaker validation set (AVA)

Figure 3. Detection accuracy on the AVA and Noise AVA validation set.

### 4.2 Comparison against the other works

For a fair comparison, we compared the proposed model against the best baseline model [2] and Chong H. and Kazuhito K. [5] work evaluated on the AVA-ActiveSpeaker dataset. Because, unlike other works, these models did not leverage additional postprocessing or preprocessing tasks, different ensemble networks, additional inputs like optical flow, and a pre-trained model. Further, these works reported their model prediction using the same metrics we used. The previous works' reported detection result, shown in Table 2, is copied from [5]. As shown in Table 2, both previous work models employed the MobileNet [6] to extract features from the audio and video inputs on relatively larger input sizes. However, our model only employed VGG-M, which is less in size and depth than MobileNet on only the video input and directly inputs the audio features to reduce the computational burden to train and extract features from audio features. Though our models input small-scale facial images and audio size, the experiment indicates that our model attained the same detection result as the state-of-the-art and baseline models.

Table 2. The comparison with other works.

| Model | Video input size | Audio input size | Visual Stream | Audio Stream | Inf. time | Auc. |
|---|---|---|---|---|---|---|
| Baseline [2] | 128x128x5(gray scale) | 64x48x1 | MobileNet | MFCC+MobileNet | - | 88.74 |
| Chong H.[5] | 128x128x10(gray scale) | 64x48x1 | MobileNet | MFCC+MobileNet | - | 89.14 |
| **Ours(M3)** | 100x100x5(gray scale) | 13x20x1 | | MFCC+VGG-M | 47.44ms | 88.929 |
| **Ours(M1)** | | | | MFCC | 44.41ms | 87.940 |

## 5. CONCLUSION

Previous active speaker detection works have not explored fusing raw MFCC audio features with ConvNet features extracted from raw images. Further, the previous works focus on improving the detection result of their proposed model, proposing a complex network with a large input size, and the inference time of their model was not reported. This work proposes a novel and simple end-to-end active speaker detection framework fusing the ConvNet feature extracted from the face regions and audio future extracted by MFCC filters. The experiments show that with more minor complications, the proposed model attained the same detection accuracy as models that employed VGG-M on the top of the MFCC. We also reported the proposed model's inference time, and our model predicts five frames within 44.41ms. With this, one can use it for real-time applications. Furthermore, the experiment shows that our MFCC-based model showed more robustness towards the noisy signal than VGG-M based audio feature extractor model. Our best-performing model also attained nearly the same detection accuracy as the state-of-the-art work. There is still much room to improve our works using more sophisticated audio features.